# Patterning of two-dimensional electron systems in SrTiO$_3$ based heterostructures using a CeO$_2$ template

D. Fuchs,[1] K. Wolff,[1] R. Schäfer,[1] R. Thelen,[2] M. Le Tacon[1] and R. Schneider[1]

[1]Karlsruher Institut für Technologie, Institut für Festkörperphysik, D-76021 Karlsruhe, Germany

[2]Karlsruher Institut für Technologie, Institut für Mikrostrukturtechnik, D-76021 Karlsruhe, Germany

Two-dimensional electron systems found at the interface of SrTiO$_3$ based oxide heterostructures often display anisotropic electric transport whose origin is currently under debate. To characterize transport along specific crystallographic directions we developed a hard-mask patterning routine based on an amorphous CeO$_2$ template layer. The technique allows preparing well defined microbridges by conventional ultraviolet photolithography which – in comparison to standard techniques such as ion- or wet-chemical etching - does not show any degradation of interfacial conductance. The patterning scheme is described in detail and the successful production of microbridges based on amorphous Al$_2$O$_3$-SrTiO$_3$ heterostructures is demonstrated. Significant anisotropic transport is observed for $T < 30$ K which is mainly related to impurity/defect scattering of charge carriers in these heterostructures.

**I. INTRODUCTION**

The discovery of two-dimensional (2D) superconductivity at the interface of oxide heterostructures, whose building blocks consist of insulators, has attracted huge fascination [1-3]. The most prominent example for this is LaAlO$_3$/SrTiO$_3$ (LAO/STO), displaying not only superconductivity below $T_c = 300$ mK, but also multiple quantum criticality [4], magnetism [5], and tunable spin-orbit coupling as well [6]. These appealing properties put LAO/STO to a canonical system for studying the impact of electronic correlations in 2D. The physics of the confined *d*-orbital states is far richer and more complex than encountered in the case of *p*-states of electron gases in semiconductors and are more compatible with a correlated 2D electron liquid (2DEL) than a 2D electron gas (2DEG) [7]. In addition, electronic transport is often found to be anisotropic. Nonlocal resistance has been observed in 2D interfacial electron systems such as LaTiO$_3$/STO [4] or amorphous Al$_2$O$_3$/STO (*a*-Al$_2$O$_3$/STO) heterostructures [8] and seems to be not unique to the 2DEL in LAO/STO.

Mesoscopic inhomogeneities could be indeed verified by measurements of the superfluid density [9] and magnetism [10,11], indicating embedded superconducting "puddles" in a (weakly localized) metallic background. Furthermore, current distribution [12] and surface potential [13] display a striped, filamentary electronic structure, too. Impurities and defects, or a net surface charge at the step edges [14] are most likely the main source for the measured anisotropic transport. However, a negative compressibility of the 2DEL [10] may also hint at an intrinsic mechanism that results in charge segregation and electronic phase separation, even in a perfectly clean and homogeneous system.

To characterize electrical transport along specific crystallographic directions resistance microbridges and Hall bars with corresponding alignment have to be patterned. Patterning techniques such as the deposition of an amorphous LAO inhibit layer (previously suggested) or ion-etching are meanwhile known to result in parasitic conductance of the etched surface and are hence not applicable here [15,16]. Wet-chemical etching is problematic as well since suitable etchants are difficult to find. In the following we describe a suitable scheme to pattern the 2DEL at STO-based heterostructures. The patterning scheme is based on an amorphous $CeO_2$ hard mask. The insulating $CeO_2$ template layer does not lead to any interfacial conductance between $a$-$Al_2O_3$ or STO and is therefore suitable to act as a hard mask. The process does not need any wet-chemical or ion-etching which conserves the $TiO_2$-terminated STO surface - necessary to obtain interfacial conductance - to largest extend.

## II. EXPERIMENTAL AND RESULTS

In the following section we describe the sample preparation and the patterning scheme, see Fig. 1, in detail. As starting material we use one side polished (001) oriented STO substrates with sizes $5\times5\times1$ mm$^3$ from CrysTec company (Berlin/Germany). To achieve interfacial conductance in $a$-$Al_2O_3$/STO-heterostructures, a single $TiO_2$ termination of the STO substrate surface is necessary. Reproducible results are obtained using similar recipes as described in literature [17]. First, the substrates were bathed in bi-distilled water for about 10 minutes, than etched in buffered ammonium fluorid (BHF) for 30s followed by a stopping bath in bi-distilled water for 10s and blow-dried with nitrogen. We did not handle the substrates in ultrasonic bath. To recrystallize the substrate surface, annealing in flowing oxygen for 5h at 950°C in a tube furnace was carried out. Before further processing, substrate surface is checked by atomic force microscopy (AFM). The single-type $TiO_2$ surface termination is verified by formation of terraces with step-heights of one STO unit cell. Step-bunching usually does not occur. The substrate miscut angle, typically 0.1°, results in a terrace width of about 250 nm. The terrace surface usually displays constant friction signal if the AFM is operated in the friction mode – demonstrating the single –type termination of the surface.

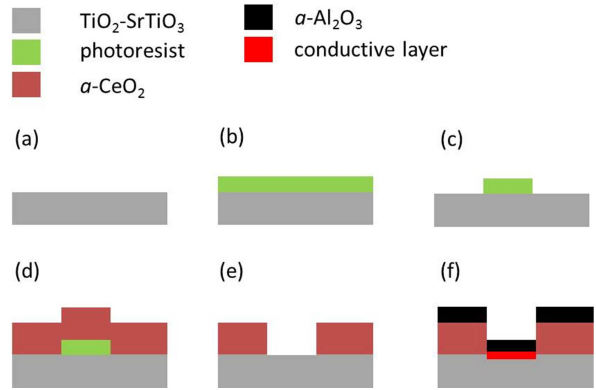

FIG. 1. Schematic illustration of the patterning process (cross section). (a) Bare $TiO_2$-terminated (001)-oriented $SrTiO_3$ substrate. (b) Substrate coated with positive photoresist. (c) Structure after UV-light exposure and development. (d) Deposition of $a$-$CeO_2$ by PLD. (e) Removing photoresist by lift-off process. (f) Deposition of $a$-$Al_2O_3$ by PLD. A conductive layer (2DEL) is formed at the interface.

After $TiO_2$ termination the structure layout of the chromium mask, see Fig. 2 (a), was reproduced on the substrate surface using positive photoresist (AZ MIR 701, Micro Chemicals). Spinning at 66 rounds per second for 60s results in a photoresist-thickness of 900 nm, which subsequently has been soft baked for 60s on a hot plate (90°C), exposed to ultraviolet light with an energy dose of 80mJ/cm$^2$ and developed for 45s in AZ 726 MiF (Micro Chemicals).

After development of the resist, the substrate was transferred to the PLD chamber to produce the $CeO_2$ template. There are also reports where AlN has been used as a hard mask [18]. The $CeO_2$ was deposited at room temperature and at an oxygen partial pressure of $p(O_2) = 0.1$ mbar from a polycrystalline $CeO_2$ target. The laser fluence was about 1.5 J/cm$^2$ resulting in a deposition rate of 0.5 Å per laser pulse. The thickness of the amorphous $CeO_2$ layer was about 75 nm.

After $CeO_2$ deposition we carried out a lift-off process using TechniStrip P1316 (Micro Chemicals) to remove the photoresist. That step results in a well-defined and sharp template hard mask, see Fig. 2 (b+c). The AFM images in Fig. 2 (d+e) demonstrate the intact surface topography, i. e., stepped surface due to the $TiO_2$ termination, of the conduction path areas. The step edges are always oriented along the same direction throughout the substrate surface. In order to produce the 2DEL the sample was again transferred to the PLD stage and an amorphous 15 nm thick layer of $Al_2O_3$ was deposited onto the $CeO_2$/STO sample. The deposition was carried out at a substrate temperature of 250°C and $p(O_2) = 10^{-6}$ mbar. The laser fluence was 1.5 J/cm$^2$ which resulted in a deposition rate of 0.1 Å per laser pulse.

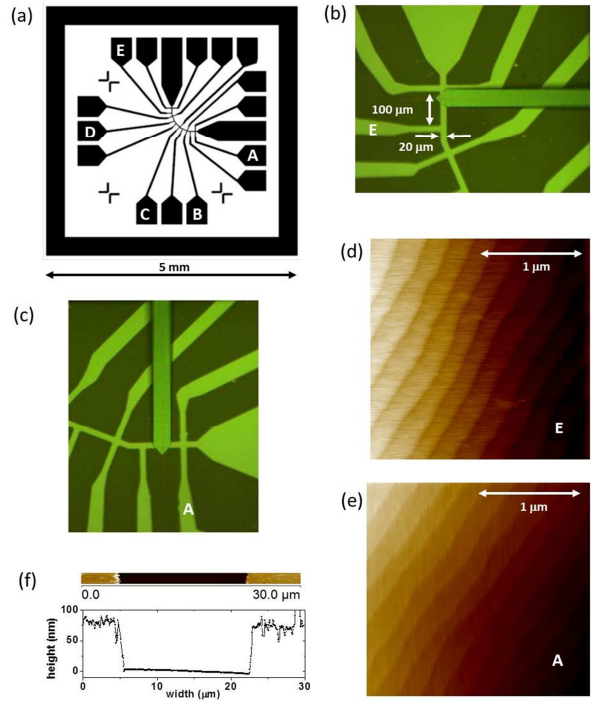

FIG. 2. (a) Mask layout (5×5 mm$^2$) for the Hall-bar arrangement. Micro bridges with a width of 20 μm and length of 100 μm are aligned at φ = 0°, 22.5°, 45°, 67.5°, and 90° with respect to the edge of substrate, i.e, the [100] direction and are labeled from A to E, respectively. (b) and (c) Micrograph of the $CeO_2$ template (dark green parts) with zoom to bridge E and A, respectively. The image was taken during AFM measurements displaying the tip as well. (d) and (e) AFM height signal taken from contact measurements. The stepped surface of $TiO_2$-terminated STO strip lines (bright green parts) is still well maintained after the production of the hard mask. Step edges are always oriented along the same direction. (f) AFM measurement across a strip line demonstrating the sharp bottom to top profile of the template structure.

Contacts to the conducting interface were provided by ultrasonic bonding through the insulting $Al_2O_3$ top layers using Al-wire. Sharp contrast between $CeO_2$ template and $a$-$Al_2O_3$/STO, see Fig. 3, enables easy identifying and contacting sample contact areas.

In Fig. 3 we show 4-point measurements of the sheet resistance $R_s$ versus temperature $T$ for the microbridges A, C, and E. Here, we have used a sample where interfacial step edges were oriented parallel to bridge E. The measurements were carried out on a physical property measurement system (PPMS) from Quantum Design using an $ac$-current of 3 μA. Distinct anisotropic resistance is found below about 30K, with the highest $R_s$ along C and the lowest $R_s$ along E. For $T > 100$ K anisotropic behavior vanishes and $R_s$ obtained for different bridges coincides. This demonstrates good homogenous conductance and reproducibility of the micro bridges by the used lift-off hard mask technique. In the temperature range where electric transport is found to be isotropic, $R_s$ is well comparable to that of unpatterned samples. Therefore, the proposed patterning technique obviously does not seem to affect conductance and is well suitable for a directional characterization of electric transport in 2DEL of STO based heterostructures.

In order to analyze the anisotropic behavior in more detail, we have fitted $R_s(T)$ in the temperature range 100 K > $T$ > 12 K assuming dominant contributions from impurity and electron-phonon scattering. Following Matthiessen´s rule, in the discussed temperature range $R_s$ is given by: $R_s = R_{imp} + R_{el-ph}$ which results in [19]:

$$R_s = A \times \left(1 - \exp(-\frac{T_A}{T})\right) \times \left[\frac{T_1}{2} \times \coth\left(\frac{T_1}{2T}\right) - T_0\right]^2 + B \times T^2 \qquad (1)$$

The constants $A$, $T_A$, $T_1$, and $T_0$ are parameters describing the dielectric permittivity $\varepsilon(T)$ of STO, and the constant $B$ characterizes the electron-phonon scattering. Fits to the data are shown by solid lines. From the obtained fitting parameters we reconstructed $R_s(T)$ also for 12 K > $T$ > 2 K, see dashed line in Fig. 3. Obviously, below 12 K $R_s(T)$ seems to consist of additional contributions besides impurity/defect scattering. However, the ratio between the largest and the smallest resistance, i. e., $R_s$(bridge C)/ $R_s$(bridge E) ≈ 1.4, at $T = 4$ K, is well comparable to the ratio which is obtained from the fits, i. e., $R_s$(bridge C)/ $R_s$(bridge E) ≈ 1.5, where $R_s$ is only dominated by impurity scattering. Therefore, the main part of the anisotropic transport is caused by impurity/defect scattering. Additional contributions by, e. g., quantum effects such as weak antilocalization (WAL) or electron-electron interaction (EEI) cannot be excluded and may be present as well but must be minor compared to impurity scattering.

Intense impurity or defect scattering may occur, on the one side, in the bulk. Flame fusion (Verneuil) -grown STO single crystals are well known to display high dislocation densities (> $10^6$ cm$^{-2}$) [20]. Most prominent are <110> dislocations with preferential {1-10} slip planes leading for, e. g., to an atypical mechanical (plastic) behavior [21]. Such <110> dislocations may also cause charge carrier scattering and therefore increased resistance perpendicular to {110} planes. This may explain why bridge C displays higher $R_s$ compared to bridge A and E.

On the other side, especially for interfacial metallic systems, defect scattering at the interface has to be taken into account as well. Brinks and coworkers [22] have shown that interfacial steps likely lead to decreased charge carrier mobility and hence increased low temperature resistance in LAO/STO heterostructures [22]. We likewise observe a small increase of the residual resistance of bridge A, which is running perpendicular to the interfacial steps in comparison to bridge E. As discussed in Ref. [18], interfacial steps may also result in further break up of inversion symmetry within the film plane resulting in spin-orbit coupling (SOC) in addition to that which usually results from symmetry breaking perpendicular to the interface. In order to investigate possible influence of SOC on the anisotropic behavior of $R_s$ we intend to carry out measurements of the magnetoresistance with magnetic field $B$ applied perpendicular to the sample surface. Besides classical Lorentz–scattering caused by impurities/defects, quantum effects such as WAL or EEI may influence magnetotransport significantly and hence may allow to reveal additional contributions to anisotropic transport.

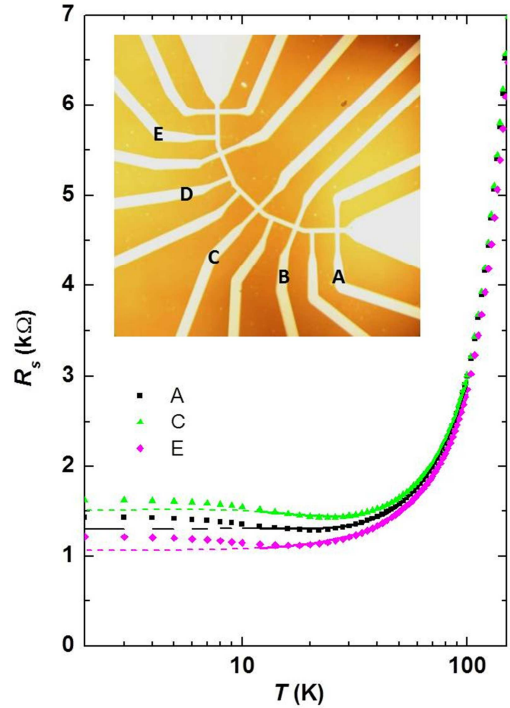

FIG. 3. Sheet resistance $R_s$ versus $T$ as obtained from 4 point measurements on bridge A, C, and E. Fits to the data according to equation (1) in the temperature range 100 K > $T$ > 12 K are shown by solid lines. Dashed lines indicate $R_s(T)$ for $T$ < 12 K, as deduced from the fitting parameters obtained before. The inset displays micrograph of the sample. Micro bridges are indicated. Sharp contrast between $CeO_2$ template (dark) and $a$-$Al_2O_3$/STO (bright) enables easy identifying and contacting sample contact pads. Interfacial step edges are oriented parallel to bridge E.

## III. SUMMARY

In order to characterize anisotropic electrical transport in the two-dimensional electron systems of $SrTiO_3$-based heterostructures we developed a hard-mask patterning routine based on an amorphous $CeO_2$ template layer produced by lift-off process. The technique allows the preparation of well-defined microbridges without applying ion-beam irradiation or etchants to the delicate $TiO_2$-terminated $SrTiO_3$ surface. The proposed patterning scheme displays good reproducibility and does not seem to influence conductance making the technique well suitable for a directional characterization of electric transport in two-dimensional electron systems of $SrTiO_3$-based heterostructures. $a$-$Al_2O_3$/STO heterostructures display anisotropic transport below about 30 K. The anisotropy in $R_s$ is dominantly related to impurity/defect scattering and likely caused by <110> dislocations in STO and step edges at the interface.

## ACKNOWLEDGEMENT

We are grateful to the Karlsruhe Nano Micro Facility (KNMF) for technical support.